\documentclass[11pt,a4paper]{article}

\usepackage{amssymb,amsmath}
\usepackage{graphicx}
\usepackage{epsfig}
\usepackage[usenames,dvips]{color}
\usepackage[usenames,dvipsnames]{xcolor}
\usepackage{hyperref}

\begin{document}

\begin{titlepage}

\begin{center}

\vskip 1.5 cm {\Large \bf Maximum Entropy deconvolution of Primordial Power Spectrum \\[4 mm]}
\vskip 1 cm {Gaurav Goswami\footnote{gaurav@iucaa.ernet.in}
 and Jayanti Prasad \footnote{jayanti@iucaa.ernet.in}\\
}
{\vskip 0.75cm IUCAA, Post Bag 4, Ganeshkhind, Pune-411007, India}

\end{center}

\vskip  .25 cm

\begin{abstract}
\baselineskip=16pt
\noindent
It is well known that CMB temperature anisotropies and polarization 
can be used to probe the metric perturbations in the early universe.
Presently, there exist neither any observational detection of tensor modes of primordial metric 
perturbations nor of primordial non-Gaussianity. In such a scenario, primordial power spectrum 
of scalar metric perturbations is the only correlation function of metric perturbations 
(presumably generated during inflation) whose effects can be directly probed through various observations. 
To explore the possibility of any deviations from the simplest picture of the era of cosmic 
inflation in the early universe, it thus becomes extremely important to uncover the amplitude 
and shape of this (only available) correlation sufficiently well. 
In the present work, we attempt to reconstruct the primordial power spectrum of scalar 
metric perturbations using the binned (uncorrelated) CMB temperature anisotropies data using 
the Maximum Entropy Method (MEM) to solve the corresponding inverse problem. 
Our analysis shows that, given the current CMB data, there are no convincing reasons to believe 
that the primordial power spectrum of scalar metric perturbations has any significant features. 
\end{abstract}

\end{titlepage}

\tableofcontents

\section{Introduction} {\label{intro}}

Observations of Cosmic Microwave Background (CMB) temperature anisotropies as well as polarization \cite{WMAP}
can be used to uncover the physics of the early universe e.g. of cosmic inflation \cite{inflation,fluctuations,infrev}. 
However, calculations of power spectra of CMB anisotropies and polarization \cite{cmbf,camb}
involve making a number of assumptions 
e.g. about the reionization history of the universe, the equation of state of dark energy etc. 
It is also usually assumed that the primordial power spectrum of scalar metric 
perturbations (denoted by sPPS in this work) is a power law (with a small running). 
One can then use the CMB observational data to put constraints on the values of various cosmological parameters \cite{COSMOMC}
including the ones specifying sPPS (usually denoted by 
$A_S$, $n_S$ etc). Since this procedure leads to \textquotedblleft reasonable\textquotedblright ~ values of these parameters, it 
is often said that a power law sPPS is consistent with the observed data. But it is worth noticing that this is just 
an assumption.

Cosmic inflation is the most actively investigated paradigm for explaining the 
origin of anisotropies in CMB sky as well as the large scale structure of the universe. 
The simplest versions \cite{inflation,fluctuations,infrev} of 
inflationary models give a smooth, nearly scale-invariant (tilted red) sPPS. But there are other models which are 
capable of giving more complicated forms of sPPS (abnormal initial conditions, multifield models, interruptions 
to slow roll evolution, phase transition during inflation, see e.g. \cite{PI1,PI2, cusps, nonTevol2, Leach1, Staro1992}). 
Are these models ruled out by the present data? 
Thus, even though power law sPPS is consistent with the data, the assumption of a power law PPS (with small running) 
is just that: a well motivated assumption. It is worth checking, how the models in which sPPS is not just a simple power 
law with a small running fare against the present available data. 

This can be done in various ways: e.g. one could try to redo cosmological parameter estimation with the actual form of 
sPPS left free (see e.g. \cite{JPPSO}). Another option is to work with inflationary models which lead to features in sPPS 
and redoing parameter estimation for those models (see e.g. \cite{DKH, PI1}). 
This exercise illustrates that (i) models in which sPPS is not this simple also do fit the data,
(ii) very often, with these models, one can get a better fit to data than power law with small running.

Given this situation, a reasonable possibility is to try to directly deconvolve sPPS from observed CMB anisotropies 
( i.e. $C_{\ell}$s).
Previous attempts ~\cite{ppsfeatures} at doing so seem to suggest the existence of features in sPPS (the statistical 
significance of which is still being assessed~\cite{armantarundec09}), e.g., a sharp infrared cutoff on the horizon scale, a 
bump (i.e. a localized excess just above the cut off) and a ringing (i.e. a damped oscillatory feature after the infrared break). 
This is consistent with many existing models of inflation and this has also motivated theorists to build models of inflation 
that can give large and peculiar features in primordial power spectrum (see \cite{PI1, PI2, cusps, nonTevol2, Leach1, Staro1992}).

Given the fact that primordial power spectrum of scalar metric perturbations is the only cosmological correlation whose effect is, 
at this stage, observable in the universe (Primordial non Gaussianity is yet to be detected in CMB data, so are B modes of 
polarization of CMB due to inflationary Gravitational waves), it becomes important to settle this issue of possible existence of 
features.

In the present work, we try a new method of probing the shape of primordial power spectrum: the Maximum Entropy Method (MEM).
We begin in \textsection \ref{Decov} by broadly describing the problem and its various attempted solutions. 
Then, in \textsection \ref{algo}, we describe in detail the algorithm that we have used. 
This is followed by \textsection \ref{cmbkernel} in which we apply the algorithm to binned CMB temperature anisotropies data.
We conclude in \textsection \ref{discussion} with a discussion of salient features, limitations and future prospects for the work.
In the appendix \textsection \ref{test}, we present the results of applying the method on a toy problem and in the process 
illustrate the use of the algorithm.

\section{Deconvolution Problem} \label{Decov}

\subsection{Formulation as an inverse problem}

We address the issue of reconstructing the shape of the sPPS 
by attempting to directly solve the (noisy) integral equations giving the CMB angular power spectrum using 
MEM. The observed CMB $TT$ angular power spectrum is given by (see e.g. \cite{Dodelson}):
\begin{equation} \label{fundaeqcont}
 {C_{\ell}^{TT}}_{\mathrm{obs}} = \int_0^{\infty} dk \left[\frac{4 \pi}{k} (\Delta_T(\ell,k,\eta_0))^2 \right] P_{\Phi} (k) 
~~+~~ {C_{\ell}^{TT}}_{\mathrm{noise}}
\end{equation}

\noindent here, $\ell$ is the multipole moment, $k$ is the wave number and the quantity in the square brackets is the 
radiation transfer function ($\eta_0$ denotes the value of conformal time today) and 
$P_{\Phi} (k) = \frac{k^3}{2 \pi^2} \langle | \Phi(k) |^2 \rangle$  is the power spectrum of the scalar metric perturbation in 
Newtonian gauge (often called Bardeen potential, $\Phi$). 
Assuming a given set of values of background cosmological parameters, the radiative transport kernel can be found 
(see \textsection \ref{cmbkernel}), we can then formulate the problem we are dealing with as the solution of a set of integral 
equations i.e. as an inverse problem.

The scalar primordial power spectrum is the power spectrum of comoving curvature perturbation:
\begin{equation}
 P (k) \equiv P_{\cal R} (k) = \frac{k^3}{2 \pi^2} \langle |{\cal R} (k)|^2 \rangle
\end{equation}

\noindent here ${\cal R} (k)$ is the mode function {\footnote{so it has mass dimension $-3/2$ rendering $P(k)$ dimensionless.}} 
of the comoving curvature perturbation on super-Hubble scale (when it has become frozen). For a power law sPPS,
\begin{equation}
 P(k) = A_S \cdot \left( \frac{k}{k_0}\right)^{n_S - 1}
\end{equation}

\noindent In matter dominated universe (at the time of recombination), at linear order in perturbation theory, 
$\Phi = (3/5) {\cal R}$ 
, so, for a power law PPS, ${C_{\ell}^{TT}}$ should be (in $\mu K ^2$)
\begin{equation}
 {C_{\ell}^{TT}}_{\mathrm{theory}} = {T_{\rm{CMB}}}^2 \cdot \left(\frac{3}{5}\right)^2 \cdot A_S ~
\int_0^{\infty} dk \left[\frac{4 \pi}{k} (\Delta_T(\ell,k,\eta_0))^2 \right] \left( \frac{k}{k_0}\right)^{n_S - 1} ~.
\end{equation}

\noindent We shall now replace 
${T_{\rm{CMB}}}^2 \cdot \left(\frac{3}{5}\right)^2 \cdot A_S \left( \frac{k}{k_0}\right)^{n_S - 1}$
by a general function $f(k)$ and try to find this function
$f(k)$. We thus have 

\begin{equation} 
 {C_{\ell}^{TT}}_{\mathrm{theory}} = 
\int_0^{\infty} dk \left[\frac{4 \pi}{k} (\Delta_T(\ell,k,\eta_0))^2 \right] f(k)
\end{equation}

\noindent with 
\begin{equation}
 f(k) = {T_{\rm{CMB}}}^2 \cdot \left(\frac{3}{5}\right)^2 \cdot  P(k) 
\end{equation}

\noindent and so the function $f(k)$ shall have values of the order of magnitude of $10^3$.

Given the temperature radiation transfer function ($\Delta_T(\ell,k,\eta_0)$), the theoretical $C^{TT}_{\ell}$ can be found 
from Eq [\ref{fundaeqcont}]
provided, we know the sPPS. The $\ell$ range for which we wish to evaluate the transfer function and the corresponding 
$C_{\ell}$s goes from $\ell = 2$ to $\ell = ~l_{\rm max}~ = 1500$. The typical behaviour of the function 

\begin{equation}
 G(\ell,k) = dk \frac{4 \pi}{k} (\Delta_T(l,k,\eta_0))^2
\end{equation}

\begin{figure} [htbp]
  \begin{center}
    \includegraphics[width=0.5\textwidth,angle = -90]{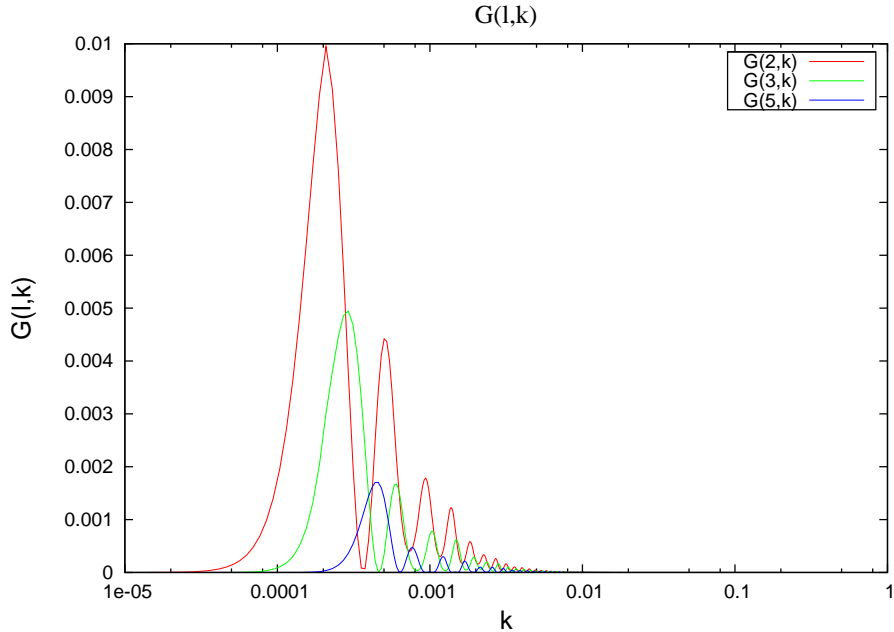}
  \end{center}
\caption{Typical behaviour of the kernel Glk for low $\ell$ values.}
\label{Glk}
\end{figure}

\noindent is shown in the Fig (\ref{Glk}) (with $dk$ chosen such that the integral in the definition of $C_{\ell}$ can be 
evaluated to a high enough accuracy). 
For every given $\ell$, the radiation transport kernel is a highly oscillatory function of the wavenumber $k$. But 
for any $\ell$, it has significant (i.e. non-negligible) values only within a small range of $k$ values. 
The brightness fluctuations roughly 
go as $j_{\ell}[k(\eta_0 - \eta_*)]$ (where $j_{\ell}$ is spherical Bessel function while $\eta_*$ is the conformal time at the epoch 
of recombination), thus the minimum value of $\ell$ sets a minimum value of $k$ at which the 
kernel takes up non-negligible values.
This procedure tells us that since the radiation transfer function is negligible for $k < k_{\rm min}$, no matter how much power is
there in sPPS at very small $k-$values, the CMB anisotropies cannot be used to probe the sPPS at these (very large scales).
This sets the $k_{\rm min}$ below which we cannot probe the sPPS. 
Similarly, given the fact that we have observations only till a maximum value of $\ell$, this sets the maximum value of $k$ 
up to which we need to sample the kernel: thus, the smallest possible angular resolution of a CMB experiment shall set the 
lmax that we can probe which shall set a $k_{\rm max}$, i.e. sPPS at scales smaller than this scale can not be probed by CMB 
experiments. 
Thus, $\ell = 2$ determines $k_{\rm min}$ while $\ell = l_{\rm max}$ determines $k_{\rm max}$. 
Within this range, one discretizes the $k-$space in such a 
way that the transfer function can be sampled sufficiently well and the above integral can be performed to the desired accuracy.
{\footnote{ As we shall see in section \ref{cmbkernel}, this number is 6200, thus our $Glk$ matrix shall have 
dimensions $1499 \times 6199 ~(~= 9292301 ~\rm{entries})$.}}

Apart from this consideration, the actual observed $C_{\ell}$s are also noisy (due to cosmic variance, instrumental noise and 
the effect of masking the sky).
Thus Eq  (\ref{fundaeqcont}) can be written as a set of linear equations
\begin{equation}
 C_\ell = \sum_{k = 1}^{n_s} G_{\ell k} f_k + C_\ell^{N}
\end{equation}

\noindent where $n_s$ is the number of bins in $k-$space and $C_\ell^{N}$ is the noise term. 
Thus the problem we wish to solve is: given the matrix $G$, the few observations ($C_{\ell}$s), the moments of the 
random variables $C_\ell^{N}$, how can we find the set of numbers $f_k$?
In this paper, we shall use the binned CMB data to find sPPS. The number of (binned and hence uncorrelated) data points 
(WMAP) is $45$ (call it $n_d$). To sample the kernel satisfactorily, we divide the $k$ space into $6200$ points 
($n_s$). Thus, we have a problem with a set of $45$ noisy linear equations and $6200$ unknowns to be determined.

\subsection{Bayesian inversion} 

Recovering the primordial power spectrum $f_k$ from the observed $C_l$ can be casted as a Bayesian inversion 
problem in the following way. The posterior probability $P(f_k|C_l,G_{lk})$ of obtaining the primordial
power spectrum $f_k$ given a kernel $G_{lk}$ and observed $C_l$ is given by:

\begin{equation}
P(f_k|C_l,G_{lk}) = \frac{P(C_l|f_k,G_{lk}) P(f_k) }{P(C_l)} 
\end{equation}
where $ P(C_l|f_k,G_{lk})$ is the likelihood and $P(f_k)$ is the prior probability. For our case the 
denominator (evidence) works just a normalization and we can ignore it.

For the case of Gaussian noise
{\footnote{
Even though the noise on $C_{\ell}$s is not Gaussian, we proceed pretending the noise to be Gaussian. This is justified because
by the central limit theorem: since the chi-squared distribution is the sum of $n_d$ independent random variables with finite mean 
and variance, it converges to a normal distribution for large $n_d$. 
}
}
the likelihood function can be written as 
\begin{equation}
 P(C_l|f_k,G_{lk}) \propto \exp[-\chi^2/2]
\end{equation}
where 
\begin{equation}
 \chi^2 =   (C_l- G_{lk} f_k)^T Cov^{-1} (C_l-G_{lk}f_k) = \sum\limits_{l=2}^{l=l_{max}} \frac{|C_l-G_{lk}f_k|^2}{\sigma_l^2}
\end{equation}
for the case when the noise covariance matrix is diagonal.

Since  for our problem the number of unknowns i.e., $f_k$ are far more than the number of known i.e., $C_ls$ 
therefore ordinary chi square minimization is of no use since it can make the chi square too low
{\footnote{even if we knew all the $n_s$ parameters, the presence of noise shall ensure that $\chi^2$ shall be a sum of $n_d$ 
normalized Gaussians.}
}
.
In order to avoid chi square taking unphysical values we need some form of regularization in the form of prior.
In place of maximizing the likelihood function we maximize the posterior probability.

It has been a common practice to consider the following form of prior for any regularization problem
\begin{equation}
P(f_k) \longrightarrow P(f_k,\lambda, S) = \exp[-\lambda S(f_k)/2]
\end{equation}
where $\lambda$ is the regularization parameter and $S$ is the regularization function. There have been 
many form of regularization function like quadratic form etc. 

In the present work we use an Entropy function $S(f_k)$ as a regularization function which is defined in the 
following way

\begin{equation}
 S(f_k) = -\sum\limits_k f_k  \left [ \ln \left( \frac{f_k}{A} \right) -1 \right] 
\end{equation}
where $A$ is a parameter which parametrizes the entropy functional we use.

With the regularization function the posterior probability distribution can be written as

\begin{equation}
 P(f_k|G_{lk},C_l) = \exp[-\chi^2/2] * \exp[-\lambda S/2] = \exp[-(\chi^2+\lambda S)] = \exp[-M(f_k)]
\end{equation}
where
\begin{equation}
M(f_k) = \frac{1}{2} \left( \chi^2 + \lambda S(f_k,A) \right)
\end{equation}

Maximum entropy method is a particular (nonlinear) inversion method. Here the regularization function $S(f_k,A)$ is 
non-quadratic so that the 
equations to be dealt with to solve the optimization problem shall turn out to be non-linear. 
%
Without such a maximum entropy (ME) constraint, the inversion problem is ill-posed (since the data can be satisfied by an infinity 
of primordial power spectra). The condition that the entropy be a maximum selects one among these.
There exist, in the literature, various arguments justifying the use of MEM over other ways of inversion (often using arguments from 
information theory
{\footnote{It is often argued that 
while the maximum likelihood method approach selects the spectrum that has the largest probability of reproducing the data, the
maximum entropy method, instead, selects the positive spectrum to which is associated the largest number of ways of
reproducing the data, i.e., the one that maximizes the information-theory definition of the entropy of the spectrum
subject to the given constraints.
}}
), at this stage, we just treat it as just another nonlinear version of the general regularization scheme. 

\section{The Cambridge Maximum Entropy Algorithm} {\label{algo}}

So, the problem that we wish to solve involves a highly under-determined system of linear equations. 
As was mentioned in the last section, one way in which we can attempt to solve this problem is to
formulate it as a problem involving the optimization of a non-quadratic function (which will require
solving a set of non-linear equations) subject to a constraint. Since the number of unknowns is so 
large, we have to solve the corresponding constrained non-linear optimization problem in a very large 
dimensional space. Also, we have other constraints that we need to take care of e.g. the components 
of $f$ are positive quantities (since $f$ is a power spectrum), so the optimization algorithm that we 
use must not cause the components of $f$ to become negative (this requirement rules out methods such 
as the steepest ascent). Similarly, since the the objective function is quite different from a pure 
quadratic form, methods such as conjugate gradient method are not very useful.

Experience has shown that one of the strategies which work (despite being complicated) is the following: 
instead of searching for a minimum in a single search direction (e.g. in steepest ascent method), 
one searches in a small- (typically three-) dimensional subspace. This subspace is spanned by vectors 
that are calculated at each point in such a way as to avoid directions leading to negative values.
The algorithm that we use is based on the one developed by Skilling and Bryan  \cite{SB1984,SB_others}
and is sometimes referred to as The Cambridge Maximum Entropy Algorithm.
It has been extensively used in not only radio astronomy but also in other fields. Here we quickly review this 
algorithm for the sake of completeness.

\subsection{Entropy and $\chi^2$}

The problem to be solved involves finding a set of $f_k ~(k = 1, 2, \cdots, n_s)$ (with maximum entropy) from a data set 
$D_{\ell} ~(\ell = 1,2, \cdots ,n_d)$. For any $f_k$, let 

\begin{equation}
 F_{\ell} = \sum_k G_{\ell k}~ f_k
\end{equation}

\noindent We shall use the following definition of entropy (the non-linear regularization function)
\begin{equation} \label{entropy_def}
 S = - \sum_k f_k [\ln(f_k/A)-1] = - \sum_k f_k \ln(f_k/eA)
\end{equation}
 \noindent here, $A$ is a fixed number (sometimes called \textquotedblleft the default\textquotedblright) 
that sets the normalization of $f$. Notice that $S(\vec{0}) = 0, S(f_k = A) = n_s \cdot A, S(f_k = eA) = 0$. 
This gives, (since $A$ is fixed), 
\begin{eqnarray}
 \partial S/ \partial f_j = \log(A/f_j), ~~~~~~~~~~~~~~~~~ \partial^2 S/ \partial f_i \partial f_j = -\delta_{ij}/f_j ~.
\end{eqnarray}

\noindent telling us that $\partial_i S(\vec{0}) = \infty, \partial_i S(f_k = A) = 0$ and $\partial_i S(f_k = eA) = -1$. 
It is easy to see that entropy surfaces are strictly convex. 
Also, the expression for the various derivatives of the entropy 
tell us that the solution $f_i = A$ is the global maximum of entropy, this fact shall be important later. 
The measure of misfit that we shall use (in order 
to use the data) is the Chi-squared function

\begin{equation}
 C(f) = \chi^2 = \sum_{\ell} (F_{\ell} - D_{\ell})^2 / {\sigma_{\ell}}^2 
\end{equation}

\noindent from which we get, the gradient of $C$ 

\begin{equation}
 \partial C/ \partial f_j = \sum_{\ell} G_{\ell j} ~2~ (F_{\ell} - D_{\ell})/{\sigma_{\ell}}^2 
 \end{equation}
and the Hessian
\begin{equation}
 \partial^2 C/ \partial f_i \partial f_j = \sum_{\ell} G_{{\ell} j} ~ \left(\frac{2}{{\sigma_{\ell}}^2} \right) G_{{\ell} i}.
\end{equation}

\noindent For a linear experiment, the surfaces of constant chi-squared are convex ellipsoids in N-dimensional space.
The largest acceptable value for $\chi^2$ at 99 percent confidence is about $C_{\rm aim} = n_d + 3.29 \sqrt{n_d}$ 
(with $n_d$ being the number of observations), see \cite{SB1984}. As the above equations show, quantities such as gradient of $C$ and 
Hessian of $C$ can be easily evaluated (though finding the Hessian of $C$ is the one of the most computationally 
expensive tasks since the matrix $G_{\ell k}$ is $45 \times 6200$ and Hessian of $C$ shall be $6200 \times 6200$ matrix).

At every iteration, instead of searching for the maximum of $S$ and minimum of $C$ along a line, we search 
in an $n$ dimensional subspace of the parameter space. So, instead of 
\begin{equation}
 {f^i}_{({\rm new})} = f^i + x ~ e^i ~~~~~~~~~~~~~~(i=1,2,\cdots,n_s)
\end{equation}

 \noindent we shall have (with $e_\mu$ being $n$ search directions)
\begin{equation} {\label{step}}
 {f^i}_{({\rm new})} = f^i + \sum_{\mu=1}^n ~ x^\mu ~ e^i_\mu 
\end{equation}

\noindent Sufficiently near any point, every function can be approximated by a quadratic function (provided 
the higher order terms in the Taylor expansion can be ignored). So, within the subspace we shall model the 
entropy and chi-squared by
\begin{eqnarray}
 S(f + \sum xe) \cong s(x) \\
 C(f + \sum xe) \cong c(x)
\end{eqnarray}
\noindent where $s(x)$ and $c(x)$ are quadratic 
\begin{eqnarray} {\label{defs}}
 s(x) = s(0) + \sum_\mu s_\mu x^{\mu} - \sum_{\mu \nu} g_{\mu \nu} x^{\mu} x^{\nu} /2 \\
 c(x) = c(0) + \sum_\mu c_\mu x^{\mu} + \sum_{\mu \nu} h_{\mu \nu} x^{\mu} x^{\nu} /2
\end{eqnarray}

\noindent which correspond to the first three terms in the Taylor series expansion of $S(f)$ and $C(f)$. The first order term in the 
Taylor expansion of $S$ is
\begin{eqnarray*}
\sum_\mu s_\mu x^{\mu} &=& \sum_{i=1}^{N} \left( \frac{\partial S}{\partial f^i}\right) \left({f^i}_{({\rm new})} - f^i \right) \\
&=& \sum_{i=1}^{N} \left( \frac{\partial S}{\partial f^i}\right) \left( \sum_{\mu=1}^n ~ x^\mu ~ e^i_\mu \right) \\
&=& \sum_{\mu=1}^n \left( \sum_{i=1}^{N}  \frac{\partial S}{\partial f^i} e^i_\mu \right) x^\mu
\end{eqnarray*}

\noindent which tells us what $s_{\mu}$ should be. Similarly, $c_\mu$, $g_{\mu \nu}$ and $h_{\mu \nu}$ can be found:
\begin{eqnarray}
 c_{\mu} &=& \sum_i e^i_\mu \frac{\partial C}{\partial f^i} \\
 g_{\mu \nu} &=& - \sum_{i j} e^i_\mu e^j_\nu \frac{\partial^2 S}{\partial f^i \partial f^j} \\
 h_{\mu \nu} &=& \sum_{i j} e^i_\mu e^j_\nu \frac{\partial^2 C}{\partial f^i \partial f^j} 
\end{eqnarray}

\noindent Thus, if we know the basis vectors, we can find the quadratic functions $s(x)$ and $c(x)$.

Obviously, the above definitions shall not be valid to arbitrary distances from the point in question. The quadratic models 
are reliable only in the vicinity of the current $f$ where cubic and higher powers can be neglected. Thus, the step size at 
each iteration must be such that 
\begin{equation}
 |\delta f|^2 \leq {l_0}^2
\end{equation}
 
\noindent for some $l_0$. We thus need to define the concept of distance in this abstract space. Recall that this means we need to 
define a metric 
\begin{equation}
 ds^2 \equiv \sum_{i j} \bar{g}_{i j} d f^i d f^j 
\end{equation}

\noindent note that the metric $\bar{g}_{ij}$ is different from the function $g_{\mu \nu}$ defined by Eq. (\ref{defs}).  
Experience (see \cite{SB1984}) has shown that the following definition of distance works well
\begin{equation} {\label{metric}}
 \bar{g}_{ij} = \frac{\delta_{ij}}{f^i}
\end{equation}

\noindent this needs to be compared with the expression for the Hessian matrix of entropy (notice that $\bar{g}^{ij} = f^i \delta^{ij}$).
It is straightforward to show that
\begin{eqnarray} {\label{equality}}
 ds^2 = \sum_{ij} \bar{g}_{ij} df^i df^j = \sum_{\mu \nu} g_{\mu \nu} x^{\mu} x^{\nu}
\end{eqnarray}

\noindent while choosing ${l_0}^2$ to be $1/5$ of $\sum f$ works well (see \cite{SB1984}). The algorithm works in the 
following way: at every iteration, when we are at a point in the $f$ space, one considers a distance region s.t. the 
quadratic model is a good approximation in that region. We now find a subspace and within this subspace, we try to 
find the place where 
\begin{enumerate}
 \item $s(x)$ is maximum,
 \item $c(x)$ equals some $\tilde{C}_{\rm aim}$, and,
 \item the distance of this new point from the old point is smaller than $l_0$.
\end{enumerate}


\subsection{Construction of the subspace}
\noindent So, how do we decide the basis vectors which span the subspace? One of our aims is to find the maximum of entropy on 
the surface of ellipsoid corresponding to $\chi^2 = C_{\rm aim}$. So, naturally, the direction of gradient of entropy must be 
one of the basis vectors. Since the metric in the space of interest is not Cartesian, there shall be a distinction between 
contravariant and covariant components of vectors in the space. Since the ``position vector'' of any point is 
$f^i$, a contravariant vector, gradient such as $\partial S / \partial f^i$ is going to be 
a covariant vector. So the first (contravariant) basis vector is
\begin{equation} \label{vec_1}
 e_1^i = \sum_j \bar{g}^{i j} \frac{\partial S}{\partial f^j} = f^i \frac{\partial S}{\partial f^i}
\end{equation}

\noindent The meaning of this direction is easy to understand by recalling its equivalent in usual Cartesian space. In the usual 
situation, $(\vec{\nabla} T) \cdot \hat{n} dr = dT$ (i.e. if we are at any point, and we go in the direction $\hat{n}$ by a distance 
of $dr$, the change in the value of the function is $dT$). It is obvious from this expression that when $\hat n$ is parallel to 
the direction of gradient, the change $df$ is maximum. Thus, to maximize the change in $f$, we shall move in the direction parallel 
to $\vec{\nabla} T$ so that 
\begin{equation}
 n^i = \sum_j  {\delta}^{i j} \frac{\partial T}{\partial x^j}
\end{equation}

\noindent This equation should be compared with the definition of the first basis vector, Eq [\ref{vec_1}]
(and since the Kronecker delta is the metric in a Cartesian space, the two equations are equivalent). 
Thus, the first basis vector tells us the direction 
in which the entropy change per unit distance is maximum. 

Similarly, another basis vector could be
\begin{equation}
 e_2^i = \sum_j \bar{g}^{i j} \frac{\partial C}{\partial f^j} = f^i \frac{\partial C}{\partial f^i}
\end{equation}
 
\noindent since we wish to change the $\chi^2$ at every iteration so that we eventually reach the $\chi^2 = C_{\rm aim}$ surface.
If we find what the two search directions (defined above) become after incrementing by $x^1 e_1 + x^2 e_2$, the direction $e_1$ 
shall stay within the subspace spanned by $e_1$ and $e_2$ but the direction $e_2$ shall go out of the subspace (see \cite{SB_others}).
This suggests that we choose more basis vectors such as
\begin{eqnarray}
 e_3^i &=& f^i \sum_j e_1^j ~~ \partial^2 C/\partial f^i \partial f^j, \\
 e_4^i &=& f^i \sum_j e_2^j ~~ \partial^2 C/\partial f^i \partial f^j
\end{eqnarray}

\noindent Experience has shown that a family of three or four such search directions gives quite a robust algorithm for solving the 
problem. In our problem, we chose the third search direction to be
\begin{equation}
 e_3^i = f^i \sum_j ~~ \frac{\partial^2 C}{\partial f^i \partial f^j} \left( \frac{e_1^j}{L_s} - \frac{e_2^j}{L_c} \right)
\end{equation}

\noindent where, the following Eqs define the lengths $L_s$ and $L_c$ which are the gradient vectors
\begin{eqnarray}
 L_s = \left( \bar{g}^{~ij} ~ \frac{\partial S}{\partial x^i} ~ \frac{\partial S}{\partial x^j} \right)^{\frac{1}{2}} ,
 ~~~ L_c = \left( \bar{g}^{~ij} ~ \frac{\partial C}{\partial x^i} ~ \frac{\partial C}{\partial x^j} \right)^{\frac{1}{2}}.
\end{eqnarray}

\noindent our experience has shown that putting the factors of $L_s$ and $L_c$ in the definition of the third 
basis vector improves the speed of convergence of the answer.



\subsection{Optimization within the subspace}

Once we have found the subspace (by finding the basis vectors in the space of all $f$s), we proceed as follows: we now wish to find the 
step, the coefficients $x$ in Eq (\ref{step}). To do this, we shall solve a corresponding constrained optimization problem in the 
$n$ dimensional subspace (as was stated in the previous subsection, we worked with $n = 3$, but we shall continue to explain the 
details for a general $n$). Since the functions $s(x)$ and $c(x)$ are quadratic, the problem in the subspace is much simpler: 
it is a simple problem of quadratic programming (quadratic objective function with quadratic constraint). The only additional 
complication is that the quadratic model is not valid to arbitrary distances from the original point, so we need to satisfy 
an additional distance constraint.

Let us begin by recalling that both the matrices $g$ and $h$ are real-symmetric. Also, the matrix $g$ is positive definite. 
The reason is as follows: the way we have defined the metric on the space (see Eq (\ref{metric}) ), 
\begin{eqnarray}
 ds^2 = \sum_{ij} \bar{g}_{ij} df^i df^j = \sum_{i} \frac{(df^i)^2}{f^i}  =\sum_{\mu \nu} g_{\mu \nu} x^{\mu} x^{\nu} \ge 0
\end{eqnarray}

\noindent where in the last step we used Eq (\ref{equality}). So, it is clear that $g_{\mu \nu}$ is a positive definite 
matrix (which implies that all its eigenvalues are positive). Thus if one of the eigenvalues of 
$g_{\mu \nu}$ is a small positive number, numerical errors can cause it to become negative. 
In our implementation of the algorithm, we choose to ignore any directions which are defined by eigenvalues which are too small.
Additional simplification occurs if we simultaneously diagonalize the two 
matrices $g$ and $h$.
{\footnote{
{\bf Simultaneous diagonalization Theorem:}
If $A$ and $B$ are real symmetric matrices and $B$ is positive definite, then there exists an invertible matrix $P$ s.t. 
$P^T B P = I$ and $P^T A P$ is diagonal. The diagonal entries of $A$ are the roots of the polynomial ${\rm det}(xB-A) = 0$.
Notice that this is different from finding a basis in which both are diagonal.
Here, if $B$ is chosen to be identity matrix, then $P$ is orthogonal and $P^T A P$ is the same as $P^{-1} A P$. This leads to 
the unique diagonal representation of the matrix (with the eigenvalues as the diagonal values). Recall that the eigenvalues of a 
matrix $M$ are the solutions of the equation ${\rm det}(M - \lambda I) = 0$.
There exist stable numerical algorithms to achieve this (see e.g. page 463 of \cite{matrixcomp}) which take in the two real 
symmetric matrices $A$ and $B$ and returns the (non-singular) matrix $P$ and the diagonal matrix $P^T A P$.
}
}

After simultaneous diagonalization, within the subspace, the quadratic model functions $\tilde{S}$ and $\tilde{C}$ 
are given by
\begin{eqnarray}
 \tilde{S}(x) &=& s_0 + \sum_{\mu} s_{\mu} x_{\mu} - \frac{1}{2} \sum_{\mu} {x_{\mu}}^2 ,\\
 \tilde{C}(x) &=& C_0 + \sum_{\mu} c_{\mu} x_{\mu} + \frac{1}{2} \sum_{\mu} \gamma_{\mu} {x_{\mu}}^2 .
\end{eqnarray}

\noindent the quantities $s_{\mu}$ etc are now defined in terms of the new basis vectors (but the same old definitions). 
Since this causes the function $g_{\mu \nu}$ to become a Kronecker delta, the distance constraint Eq. looks like 
\begin{equation}
 l ^2 = \sum_{\mu} {x_{\mu}}^2 \leq {l_0}^2 ~~~ (\simeq 0.1 \sum f ~\rm{to}~ 0.5 \sum f)
\end{equation}

We chose the coefficient on the RHS to be $0.2$ and we verified that the actual value of this number is unimportant. 
Typically, the function $\tilde{C}$ is such that all its eigenvalues are (also) positive, then, the minimum value of the function 
$\tilde{C}$ in the subspace (where the above definitions work) is
\begin{equation}
 \tilde{C}_{\rm{min}} = C_0 - \frac{1}{2} \sum_{\mu} \frac{C_{\mu}^2}{\gamma_{\mu}}
\end{equation}

\noindent Thus, no matter what the global aim $C_{\rm{aim}}$ is, at a given iteration, within the subspace, we 
can not get to any values below $\tilde{C}_{\rm{min}}$. In fact, even trying to achieve $\tilde{C}_{\rm{min}}$ 
is not a great idea since in that case we shall not use any information about $\tilde{S}$. 

The real challenge in the subspace is to satisfy the distance constraint. Many different elaborate tricks have been mentioned in the 
literature to do this. We choose to not worry about getting a quick answer, hence we do the following:
in order to ensure that the distance constraint always gets satisfied (i.e. we do not go too far from the present location in just 
one step), we shall choose to have a $\tilde{C}_{\mathrm{aim}}$ which is not too different from $C_0$ (the present
value of $\chi^2$). We thus choose 
\begin{equation}
 \tilde{C}_{\rm{aim}} = {\rm{max}} (a \tilde{C}_{\rm{min}} + (1-a) C_0, C_{\rm{aim}})
\end{equation}

\noindent with $a$ chosen to be a small number (e.g. $0.01$). This causes the algorithm to take very small ``baby steps'' 
towards the answer. Numerical experience has shown that as far as our problem is concerned, this is good enough. Of course, the
actual value of $a$ or $C_{\rm{aim}}$ chosen is not important as long as the distance constraint gets satisfied. 

The problem in the sub space is thus simplified to finding the point $x_a$ such that the function $\tilde{S}$
is maximum subject to the constraint that $\tilde{C} = \tilde{C}_{\mathrm{aim}}$ (and an additional constraint that the 
distance constraint must get satisfied). The technique of Lagrange's undetermined multiplier is useful here: 
we wish to find the point on the curve $\tilde{C} = \tilde{C}_{\rm{aim}}$ where $\tilde{S}$ is maximum, to find the 
desired point, we consider the set of points at which all the partial derivatives of the function 

\begin{equation} {\label{Qeq}}
\tilde{Q} = \alpha \tilde{S} - \tilde{C}
\end{equation}

\noindent (for an undetermined $\alpha$) vanish. For any $\alpha$, such points are given by
\begin{equation}
 x_a = \frac{\alpha S_a - C_a}{\gamma_a + \alpha}
\end{equation}

\begin{figure}
  \begin{center}
    \includegraphics[width=0.5\textwidth,angle = -90]{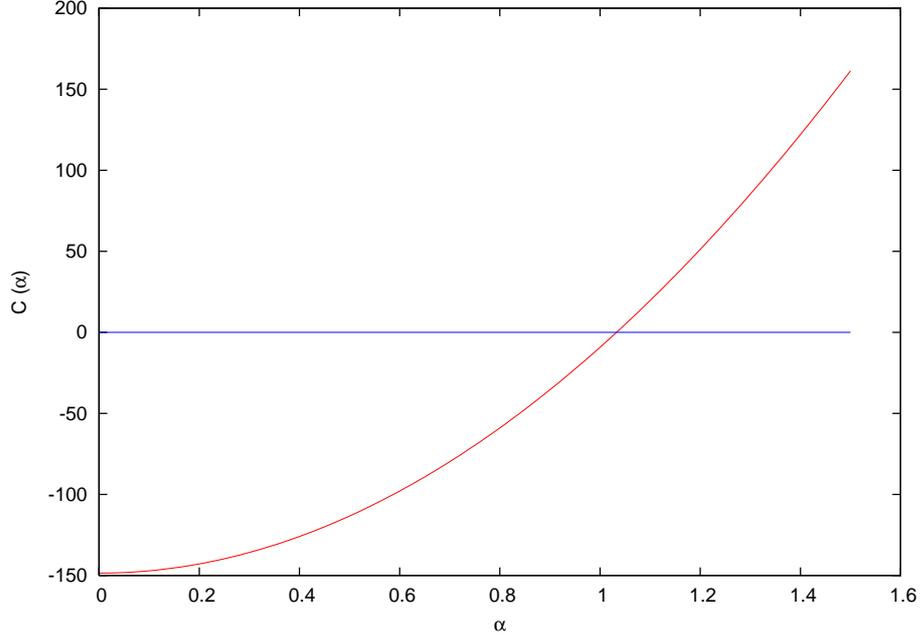}
  \end{center}
\caption{The typical behaviour of $C(\alpha) = \tilde{C}(\alpha) - \tilde{C}_{\mathrm{aim}}$ as we change $\alpha$. 
The existence of a unique solution to the Eq $C(\alpha) = 0$ is absolutely necessary for the algorithm to work. Since 
$C(\alpha)$ changes very quickly as we change $\alpha$, we need to find the root of $C(\alpha) = 0$ to a high accuracy.}
  \label{C_alpha}
\end{figure}

\noindent So, for every value of $\alpha$, find the value of $x_a$ and then the function $\tilde{C}$: 
we are after that value of $\alpha$ which leads to $\tilde{C} = \tilde{C}_{\mathrm{aim}}$ so we look for a 
solution of the equation $\tilde{C}(\alpha) = \tilde{C}_{\mathrm{aim}}$. The function $\tilde{C}(\alpha)$ is 
a monotonically increasing function of $\alpha$, see fig. (\ref{C_alpha}).
Since the function $\tilde{C}(\alpha) - \tilde{C}_{\rm{aim}}$ often happens to be a quickly changing function of $\alpha$ 
(especially while it is changing its sign), the solution for $\alpha$ needs to be found to a high tolerance level.

\subsection{Stopping criterion}

In solving the constrained optimization problem, one fact which becomes important is the following: at the point at which the 
constrained optimization problem gets solved, the gradient vectors of the two functions become parallel. Thus, if we find the 
unit vector in the direction of gradient of entropy and in the direction of gradient of chi squared function, the dot product 
of these two unit vectors (defined using the entropy metric) must become negligible as we head towards the point at which the 
constrained optimization problem gets solved. The unit vectors in the directions of gradients are (with $L_s$ and $L_c$ 
defined previously)
\begin{eqnarray}
 U^s_i = \frac{1}{L_s} \frac{\partial S}{\partial x^i}, ~~~ U^c_j = \frac{1}{L_c}\frac{\partial C}{\partial x^i}.
\end{eqnarray}

\begin{figure}
\begin{center}$
 \begin{array}{c}
\includegraphics[width=2.3in, angle = -90]{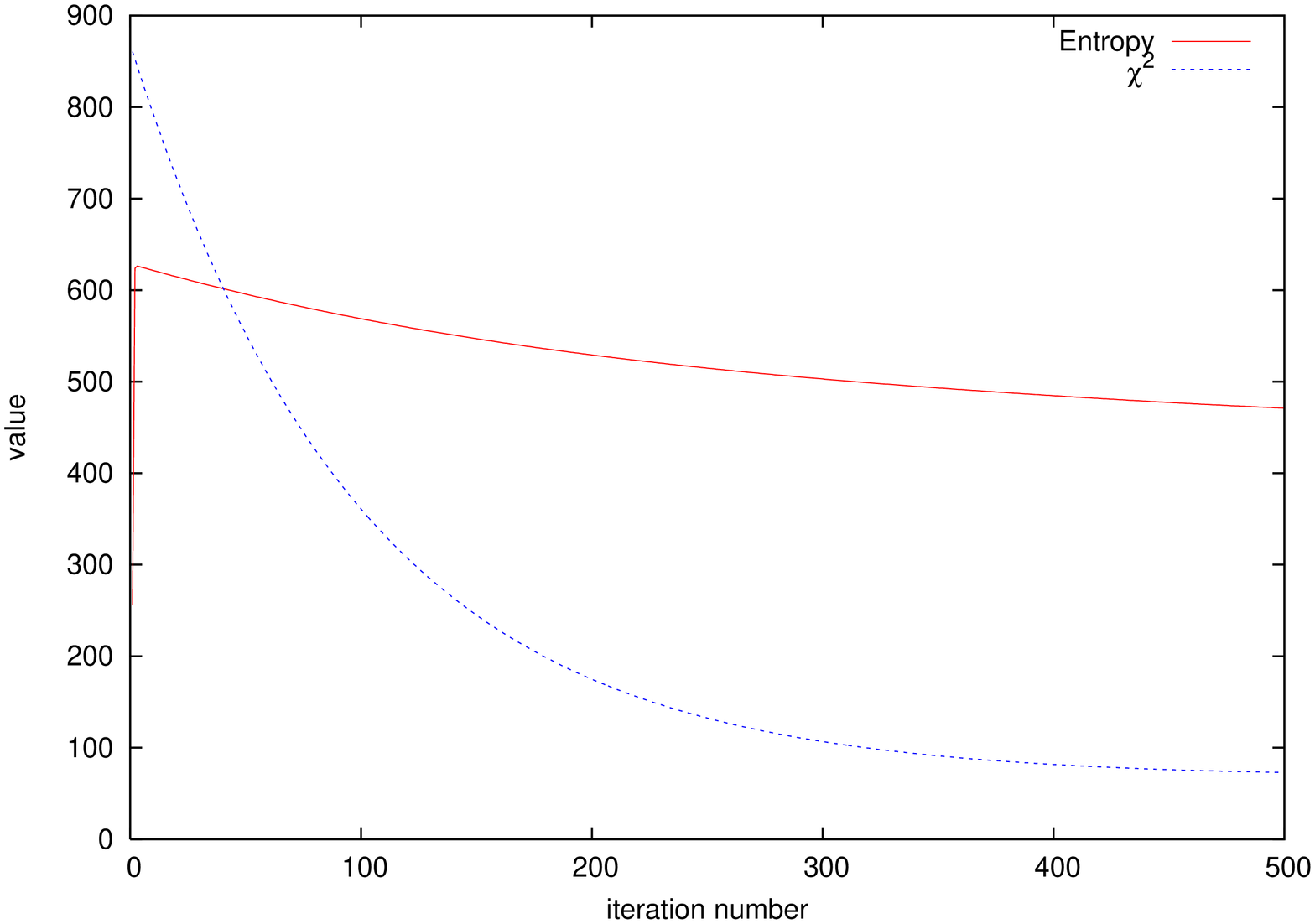} 
\includegraphics[width=2.3in, angle = -90]{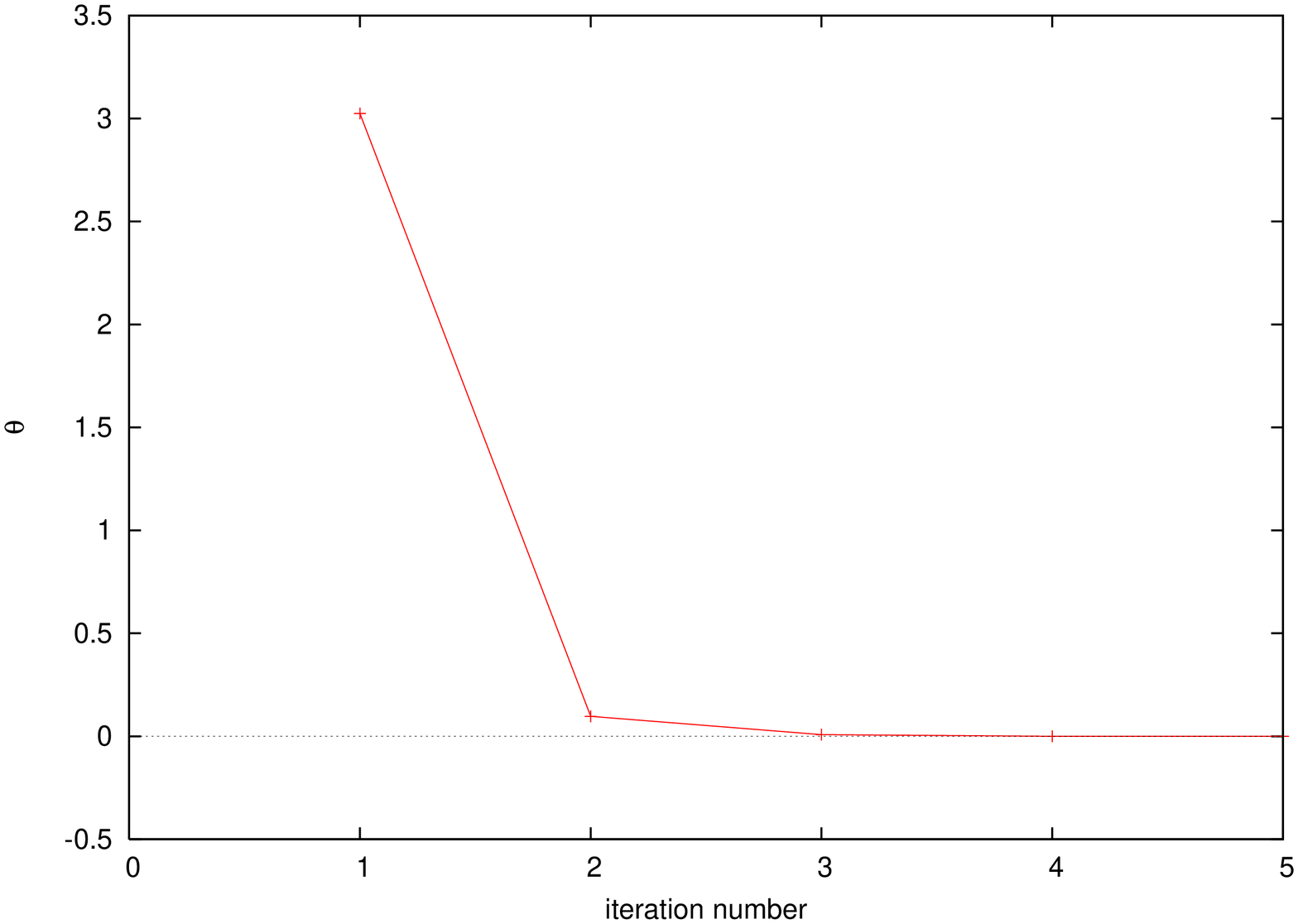}
\end{array}$
\end{center}
\caption{(i) The typical evolution of entropy and $\chi^2$ as the algorithm advances and (ii) the illustration of 
the fact that the angle $\theta$ (in radians) drops very quickly as the algorithm proceeds.}
  \label{evol}
\end{figure}

\noindent We thus expect the angle
\begin{equation}
 \theta = cos^{-1} \left( \bar{g}^{~ij} ~ U^s_{i} ~ U^c_{j}\right)
\end{equation}

\noindent to become too small (compared to a unit radian) as the algorithm proceeds (Fig(\ref{evol})).

\section{Recovering Primordial Power Spectrum} \label{cmbkernel}

In this section, we shall (i) test the formalism presented in the previous section by trying to recover a featureless as well as 
feature-full sPPS from simulated noisy CMB data and (ii) apply the algorithm to actual WMAP 7 year binned TT angular power spectrum 
{\cite{WMAP}}
to recover the Primordial Power Spectrum.
Thus, to begin with, we shall find out the radiation transfer function for the simplest set 
of assumptions, inject a featureless sPPS and get noise-free $C_{\ell}^{TT}$ (which we shall refer to as theoretical $C_{\ell}$s). 
Next we shall add noise to these pure $C_{\ell}$s. 

\subsection{The radiative transport kernel}
First, we need to set the values of the various cosmological parameters and get the corresponding 
radiation transfer function. This can be done by making use of the codes such as CMBFAST \cite{cmbf}, 
CAMB \cite{camb} or gTfast \cite{gtf}. The results in this section are got from transfer function found 
using the code gTfast which itself is based on CMBFAST (version 4.0). 
It is important to notice that since in this work we shall only use the $TT$ data, so, we only calculate 
the temperature radiation transfer function. To find out the transport kernel, 
we assume that the universe is spatially flat and dark energy is a cosmological constant (i.e. we have a 
spatially flat $\Lambda$CDM universe) and set the values of the cosmological parameters to their WMAP 
nine year values \cite{WMAP} (WMAP9 + bao + h0): 
the values of various parameters to be fed into the code gTfast are given in table \ref{Cosmo_para_1}.
We also assume that there are no tensor perturbations to the metric.  
We use Peebles recombination (rather than using RECFAST) and 
assume that the Primordial fluctuations are completely adiabatic.
Finally ,we shall not correct the transfer function for lensing of CMB, SZ effect or other 
effects that cause secondary anisotropies of CMB. 
This shall give us the radiation transfer function from which we can easily evaluate the 
matrix $G_{\ell k}$. For the case we are dealing with, the matrix $G_{\ell k}$ shall have dimensions $1500 \times 6200$.
Finally, we would like to state that the results one obtains and conclusions that one draws should better not depend on the 
exact values of these parameters. 


\begin{table} [htbp]
\centering
    \begin{tabular}{|l|l|}
        \hline
        Parameter & value \\ \hline
	lmax & 1500 \\
	ketamax & 3000 \\
        $\Omega_b$ & 0.0472 \\ 
	$\Omega_c$ & 0.2408 \\ 
	$\Omega_{\Lambda}$ & 0.712 \\ 
	$\Omega_{\nu}$ & 0.0\\
	$H_0$ & 69.33 km/s/Mpc\\
	Tcmb & 2.72548 K\\
	$Y_{\rm{He}}$ & 0.308\\
	$N_{\nu}$(massless) & 3.04\\
	$N_{\nu}$(massive) & 0.0\\
        $\tau$ & 0.088 \\
        \hline
    \end{tabular}
\caption{The values of various parameters for the run.}
\label{Cosmo_para_1}
\end{table}

\subsection{Recovering test spectra}

\begin{figure} [htbp]
   \begin{center}
    \includegraphics[height=0.5\textheight,angle = -90]{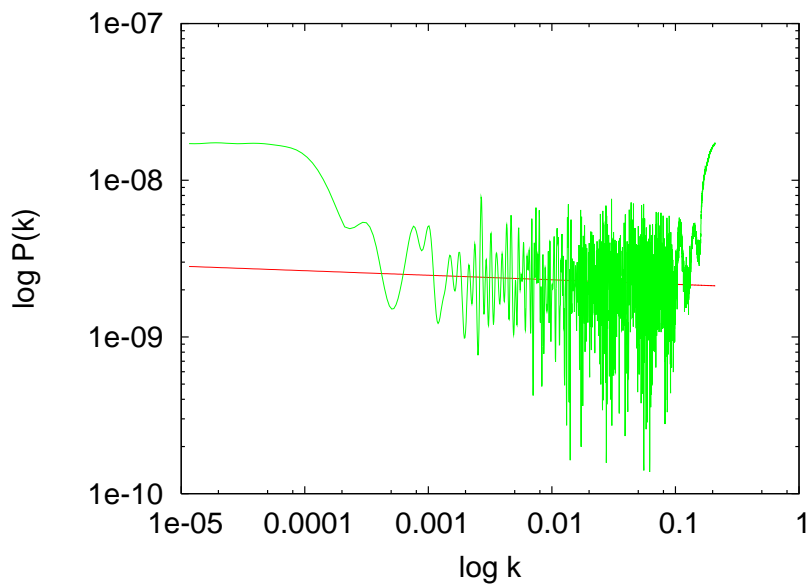}
   \end{center}
\caption{Recovery (green curve) of an injected featureless tilted red sPPS (the red line) using simulated unbinned CMB data. 
The artificially added noise is dominated by cosmic variance for small (up to 600) $\ell$ values and by instrumental noise at 
larger $\ell$ values. This result is obtained when the parameter $A$ in Eq (\ref{entropy_def}) is set to the value 
$5.4 \times 10^{4}$.}
\label{unbinned}
\end{figure}

We can now inject a test sPPS which is a power law with $A_S = 2.427 \times 10^{-9}, n_S = 0.971$
(with $k_0 = 0.002 \rm{Mpc}^{-1}$ and $T_{\rm cmb} = 2.72548 \times 10^{6} \mu$K) and 
get the corresponding theoretical $C_{\ell}$s, and add noise. The noise we add is dominated by cosmic variance at low 
$\ell$ (less than 600) values while for high $\ell$ values, the noise is dominated by instrumental errors. 
Fig (\ref{unbinned}) shows the result of using the algorithm described in the previous section to recover the sPPS 
in the present case. The following points are worth noting:
\begin{enumerate}
 \item To get the result shown in Fig (\ref{unbinned}), we set the parameter $A$ in Eq (\ref{entropy_def}) to be $5.4 \times 10^{4}$.
As was stated, the solution $f_i = A$ is the location of global maximum of entropy in the $f$ space. From 
\begin{equation}
 {T_{\rm{CMB}}}^2 \cdot \left(\frac{3}{5}\right)^2 \cdot  P(k) = f(k)
\end{equation}
 \noindent it is clear that $f = A = 54000$ corresponds to $P(k)$ being $2 \times 10^{-8}$. Thus, this value of $A$ corresponds 
to the situation in which $P(k) = 2 \times 10^{-8}$ is the solution with the maximum value of entropy.
 \item In an actual CMB experiment (such as WMAP) the amount of noise (instrumental as well as that due to cosmic variance) 
is not the same for all scales, which means that our data is not equally good for all values of $k$. 
Fig (\ref{unbinned}) shows that at scales at which the noise is large (very low and very high $\ell$ values which will 
correspond to very low and very high $k$ values), the recovered $f(k)$ tends to approach the value $A$, the recovery (at these scales) 
tends to be poor.
Thus, \emph{at scales at which the noise is too large (or the kernel takes up negligible values), 
the recovery depends on what is the prior information we have about the solution}.
Thus the range of $k$ values in which we can recover the PPS is too restricted. 
 \item Even at scales at which the noise is smaller (and at which we hope to recover well), we can have wiggly artificial features in the 
recovered PPS (in the form of peaks and dips). In the recovered power spectrum there could exist three kinds of features: 
(i) those which are actually there in the injected PPS (which are not there in the present case), 
(ii) those which are not there in the PPS but got introduced by the algorithm itself (these shall change as we change $A$) 
and finally, (iii) those which are artifacts of the added noise (a particular realization of the noise shall have outliers, 
if we consider different realizations of the noise, we shall get different recoveries).
 \item The scales at which we typically introduce features in the sPPS are roughly $10^{-3}$ MPc$^{-1}$ to $10^{-2}$ MPc$^{-1}$. 
We would like the recovery to be good at these scales. 
If we have data till very large value of $\ell$, and the noise at these large $\ell$ values is very low 
compared to the noise at $\ell$s corresponding to the above scales, the algorithm shall ignore the few data points with 
larger noise and try to only take the data at the other scales seriously. Thus, if we wish to recover better at these scales
we must focus on recovering the PPS using only the data from the $\ell$ values corresponding to these scales. Thus, 
\emph{having data till larger values of multipole moment with lesser noise may not help}.
\end{enumerate}

\begin{figure} [htbp]
  \begin{center}
    \includegraphics[height=0.5\textheight,angle = -90]{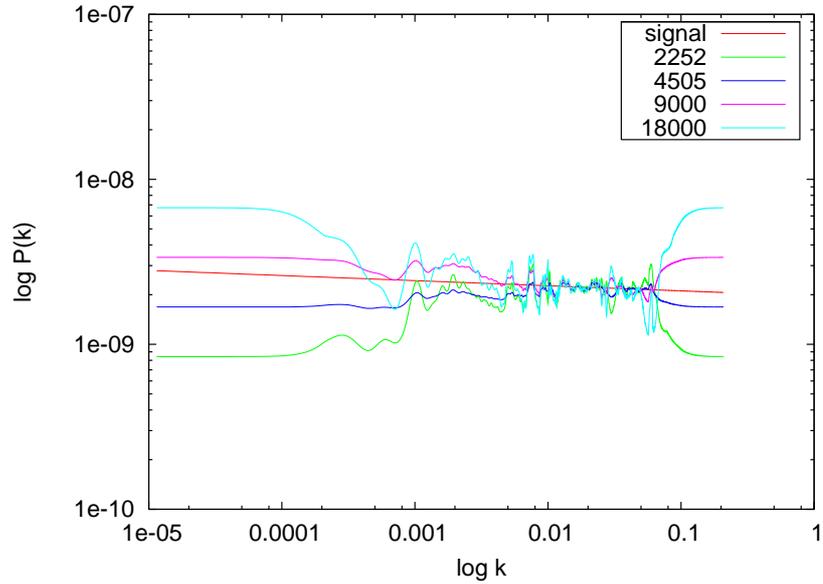}
  \end{center}
\caption{Recovery of an injected featureless tilted red sPPS using simulated binned data. The red straight line is the injected signal
while the different curves correspond to different values of $A$. Since we do not know how to fix the solution corresponding to 
Global maximum of entropy, we can not know the value of $A$.}
\label{binned_diff_A}
\end{figure}

\begin{figure} [htbp]
  \begin{center}
    \includegraphics[height=0.5\textheight,angle = -90]{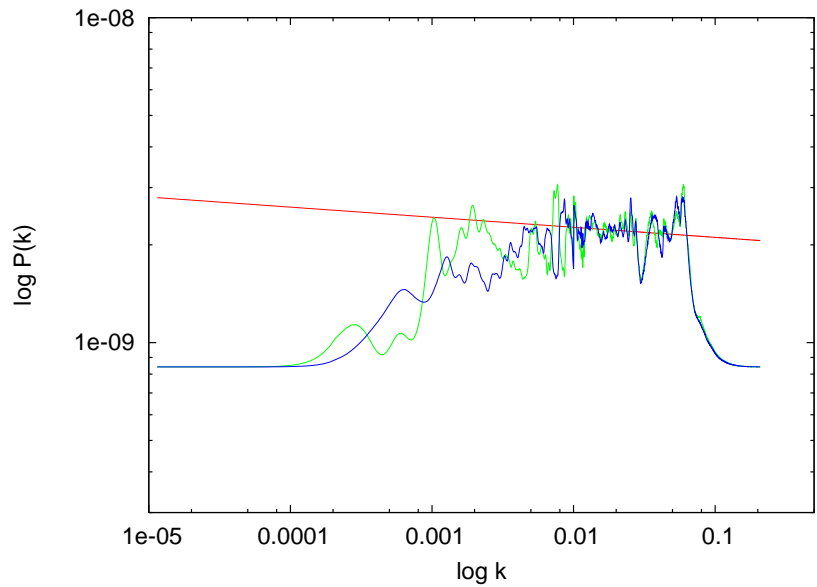}
  \end{center}
\caption{The recovery at scales at which the data has lesser noise is not given by $f_i = A$ but is dependent on the specific 
realization of the noise added. The recovery in this case is done for $A = 2252.0$.}
\label{binned_diff_realization}
\end{figure}

\begin{figure} [htbp]
  \begin{center}
    \includegraphics[height=0.5\textheight,angle = -90]{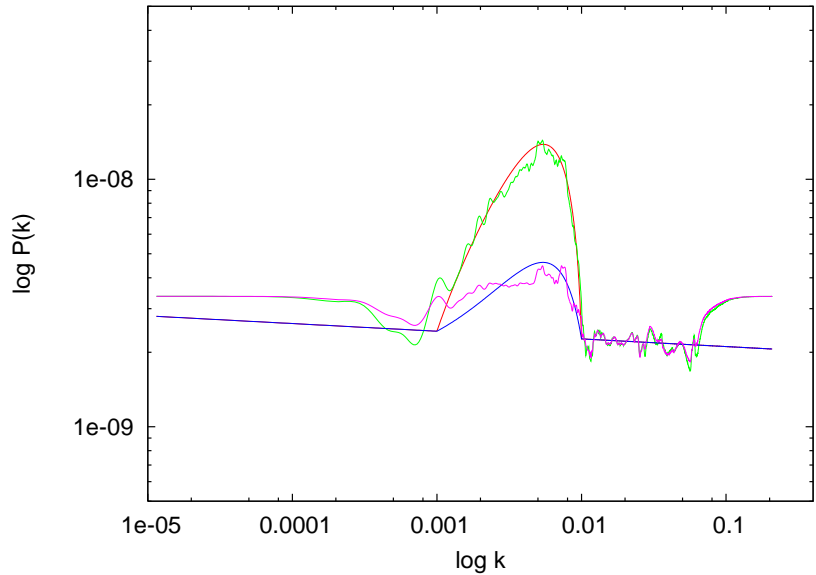}
  \end{center}
\caption{Recovery of spectrum with bumpy features. Here, $A$ is set to 9000. The red and blue curves are the injected spectra 
while green and pink ones are the recoveries. Had we introduced a feature at scales where the recovery goes back to the global 
maximum of entropy ($A$) we could not have recovered it.}
\label{binned_bumpy}
\end{figure}


In practise, the process of masking the sky causes the various $C_{\ell}$s to get correlated. The simplest situation in which we 
can hope to recover the sPPS is the one in which the the data points corresponding to different 
$\ell$ values are uncorrelated. This happens for the binned CMB data set (which has data only for 45 $\ell$ values). To make use of 
the binned data, we also work with a binned kernel which is defined
\begin{equation}
 G_{\ell k}^{\rm avg} = \sum_{\ell = \ell_{\rm min}}^{\ell_{\rm max}} \frac{G_{\ell k}}{N}
\end{equation}

\noindent where $N$ is the number of $\ell$ values in the bin. By using this averaged kernel and applying the algorithm to 
simulated binned data (with the added noise equal to the noise for WMAP 7 year binned data), we get the results shown in Fig 
(\ref{binned_diff_A}) (this time we show the results for many $A$ values). 
We again get an answer which at scales at which the noise is large, tends to the value of the default 
(i.e. $A$) while at scales at which the noise is relatively low, the recovery tends to fluctuate around the featureless 
injected signal. 
For a fixed value of $A$, the recovery at scales at which the noise is relatively lower shall be different if we consider 
different realizations of the noise. This is illustrated in Fig (\ref{binned_diff_realization}): here the recovery shall be 
the same at scales with no data and shall be different at scales with data.
The key question is whether we can recover features in the sPPS by this method. The fact that this can be done is 
illustrated in Fig (\ref{binned_bumpy}): we just introduce a bumpy feature between the scales 
$10^{-3}$ MPc$^{-1}$ to $10^{-2}$ MPc$^{-1}$ and vary its height and see that unless the height of the bump is too small, 
the algorithm can recover it. Of course if we introduce a feature at a scale at which the data is not good or at which the 
kernel takes up negligible values, the feature shall not be recovered. Moreover, it is not surprising that the recovery is 
much better if the feature is more prominent. 




\subsection{WMAP 7 year binned CMB data}

\begin{figure} [htbp] 
  \begin{center}
    \includegraphics[width=0.6\textwidth,angle = -90]{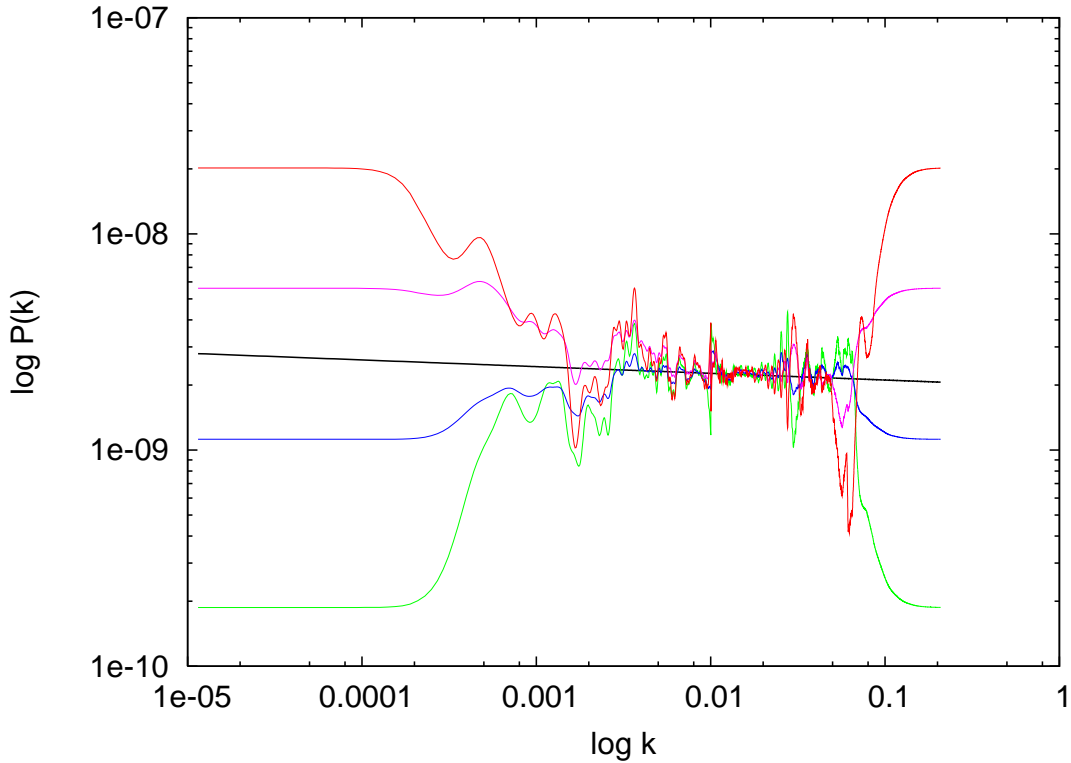}
  \end{center}
\caption{The result of applying the algorithm to binned WMAP 7 year $TT$ data. The solid black straight line corresponds to 
the Maximum Likelihood result that one gets if one assumes the sPPS to be a power law. The curves correspond to the following 
values of $A$: $A = 54000$ (red), $A = 15000$ (pink), $A = 3000$ (blue), $A = 500$ (green).}
\label{rec_binned_7}
\end{figure}

In this sub-section we apply the algorithm to actual CMB data. We use WMAP 7 year binned $TT$ data set and use it to recover the 
sPPS. The result is shown in fig \ref{rec_binned_7}.
The details of the recovery of course depend on the chosen value of the parameter $A$. 
In the present context, the value of $A$ represents our a priori knowledge (without using any data) of how much we think 
should be the scalar fluctuation in the metric in the early universe. 

\begin{figure}[hbp]  
\begin{minipage}[b]{0.45\textwidth} 
\centering \includegraphics[width = 1.9 in, angle = -90]{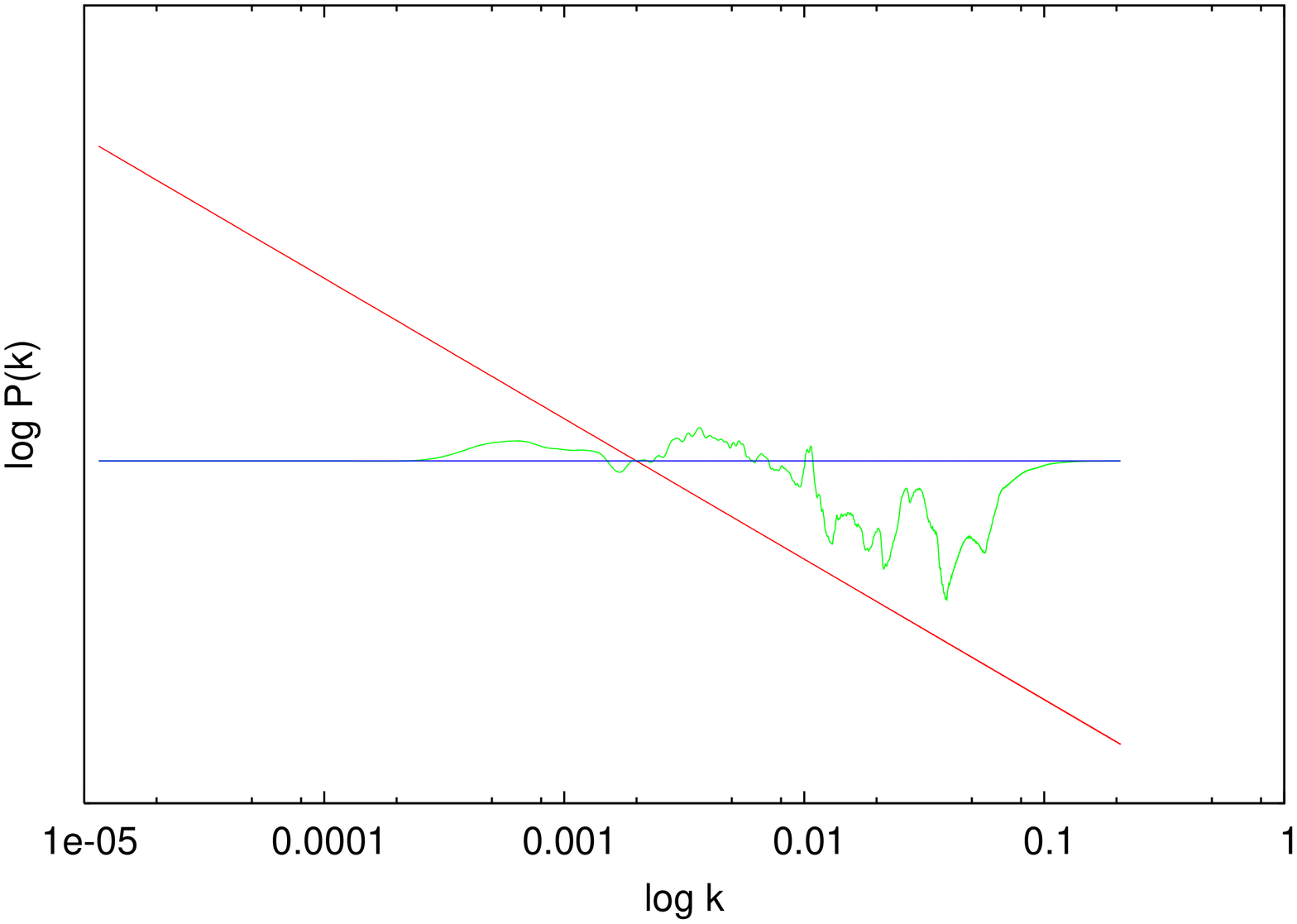} 
\caption{The red line is the WMAP ML power law sPPS. If we set $A = A_s$ (thus, the blue line is the solution corresponding to 
global maximum of entropy), we recover the green curve shown. The range of $\log P(k)$ axis is from $2.0 \times 10^{-9}$ to 
$3.0 \times 10^{-9}$.}\label{A_As}  
\end{minipage}  
\mbox{\hspace{0.5cm}} 
\begin{minipage}[b]{0.45\textwidth}  
\centering \includegraphics[width = 2.0 in, angle = -90]{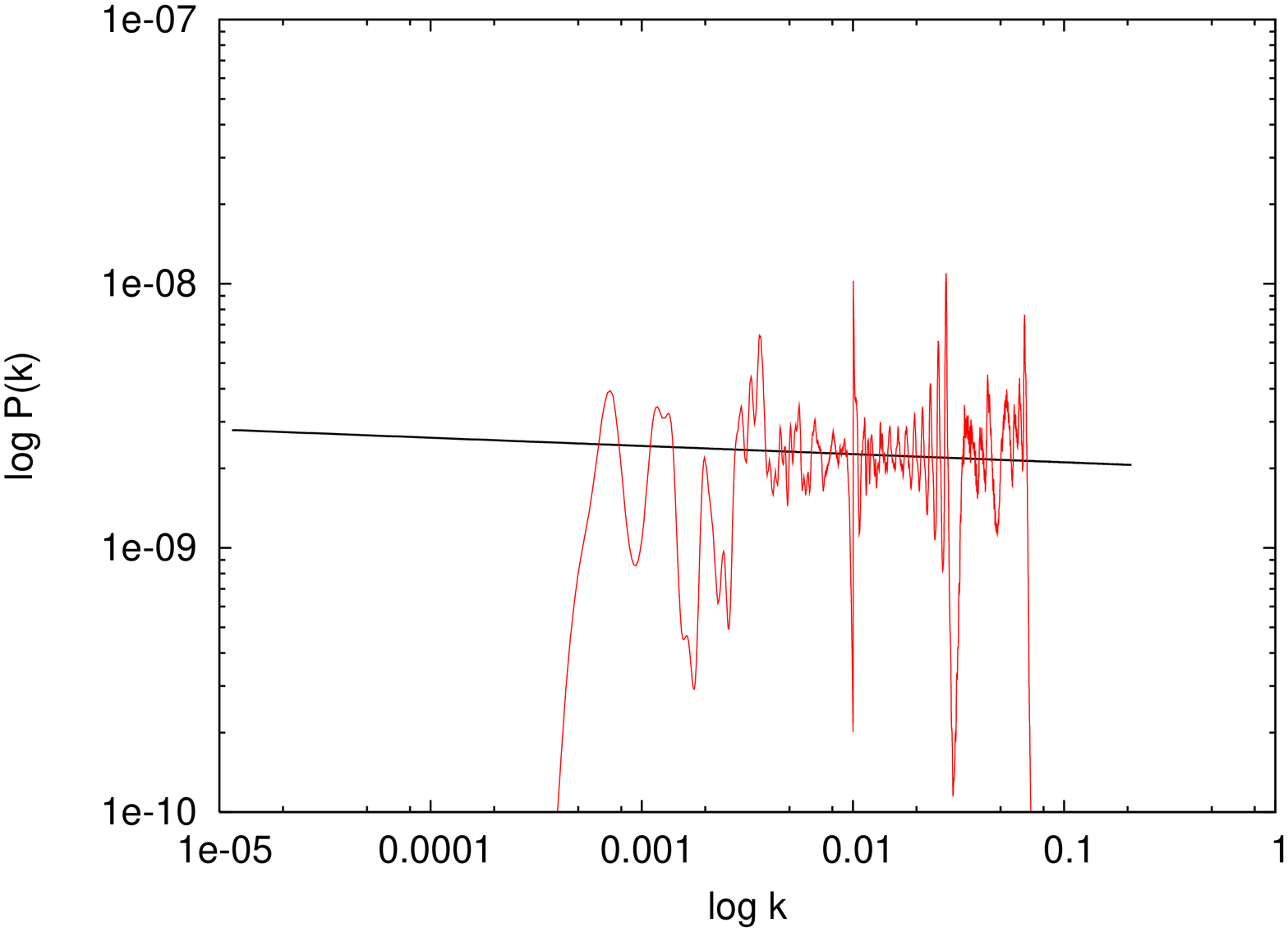} 
\caption{Choosing $A = 1.0$ causes the features to get overly \textquotedblleft stretched \textquotedblright 
and we find apparent IR and UV cut-offs in power.}  \label{rec_1}
\end{minipage} 
\end{figure}

It may appear that if the conclusion depends on such an a priori knowledge, we may not get anything worthwhile. 
But the following fact is worth noting: it is seen that at scales at which the noise is lesser, even though 
the recovered $P(k)$ depends on the value of $A$ chosen, this dependence is quite weak and quite predictable 
(as we increase $A$ a lot or decrease it a lot, the recovery just ``stretches'' in the $P(k)$ direction in 
the $\ln {P} - \ln {k}$ plane).
An interesting exercise is this: if, without using the CMB data, we still knew that the amplitude of the scalar 
metric perturbations is (roughly) $A_s$, then what can this method of deconvolution tell us about the sPPS? 
Fig (\ref{A_As}) illustrates how the red tilt of the PPS can be \emph{detected} in such a case. 
One can keep on decreasing the value of $A$ and see what happens. 
In this context, the case of $A = 1$ is very interesting since this corresponds to using another familiar definition of entropy, 
the recovery for this case is illustrated in Fig (\ref{rec_1}).
What is interesting is that if we choose $A$ to be too small, we begin to get an IR cut-off not very different from the one reported 
in the literature previously (see \cite{ppsfeatures}), but, we also get an apparent UV cut-off. Moreover, such a small value of $A$ 
causes the artificial features to get stretched so much that we may not consider the reconstruction to be trustworthy in this case.


\section{Summary and discussion} \label{discussion}

In this work, we attempted to probe the amplitude and shape of scalar primordial power spectrum (sPPS) using the CMB data. 
We fixed the values of various cosmological parameters (apart from the ones specifying the sPPS itself) and formulated the 
problem as an inverse problem. To solve the inverse problem, we use the maximum entropy method which is a non linear 
regularization method. There exist many possible ways to employ the maximum entropy regularization, we use a particular 
definition of entropy and a particular algorithm to solve the corresponding constrained non-linear optimization problem 
in a very large dimensional parameter space. 

The way we have formulated the problem, there exists a parameter (which we called $A$) whose value decides the location of 
global maximum of entropy in the space of all $P(k)$s. In the absence of any data, the algorithm shall just send every 
initial guess to the global maximum of entropy. Even in the presence of data, the following is worth noting
\begin{enumerate}
 \item at scales where 
\begin{enumerate}
 \item we have noisy data (so, little or no information), or, 
 \item the kernel (to be inverted) takes up negligible values (again too large or too small $k$ values),
\end{enumerate}

\noindent the $P(k)$ recovered by MEM depends on the value of $A$ chosen (as $P(k) = A$ is the ME solution), while 
at the scales where the data is good, we recover something which has comparatively lesser dependence on what $A$ we choose.
 \item at scales at which the data is good, the $P(k)$ recovered by MEM is consistent with a power law primordial power spectrum 
(with any possibly small deviations which we can not say anything about at this stage). This can be seen by comparing fig 
(\ref{rec_binned_7}) with figs (\ref{binned_diff_A}) and (\ref{binned_bumpy}). 
While the existence of any small deviations from power law behaviour can not be completely ruled out, this analysis 
reinforces our belief that any such possible deviations must be small. 
\end{enumerate}

This is by no means the last word on the existence of features in sPPS, this is not even the last word on the use of MEM 
for this purpose.The implementation of our algorithm to this problem till now does not seem to give any reason to believe 
that there are any serious deviations from the power law. We would like to mention that this is not completely unexpected, 
even in the light of existing papers such as \cite{ppsfeatures} because the error bars at scales at which the features were 
recovered in those works are very large: the maximum entropy method can not claim any features at scales where the error 
bars are so large {\footnote{This ``rules out'' (or at least renders them untrustworthy) 
many models of inflation considered in the literature in recent times.}}. 
This analysis shows that \emph{at scales at which the CMB data is trustworthy, the Primordial Power Spectrum of scalar metric 
perturbations is, to a very good approximation, a power law}. 

In future, one can look at the following prospects.
We should be able to solve this problem of possible existence of features in sPPS without assuming the 
values of other cosmological parameters (i.e. without formulating this problem as a simple inversion problem). 
Even in the present formulation, there may be ways of combining results from different values of $A$ to get a better 
recovery. 
One may wish to use the actual WMAP likelihood (or rather, the corresponding $\chi^2_{\rm eff}$) as a 
measure of misfit, but this is not easy in the way we have attempted to solve the problem (we need to know the $\chi^2$ and 
its first two derivatives). Also, we have lost a lot of information in the process of binning the kernel and working with 
the binned, uncorrelated data. We would like to use all that lost information. Similarly, we have only used the $TT$ angular
power spectrum of CMB, we would also like to use the polarization spectra to probe the sPPS. We may also need to post-process
the recovered sPPS to get more useful information. Another interesting possibility worth exploring is the connection of 
Maximum Entropy deconvolution with other ways of deconvolution (e.g. Richard Lucy deconvolution).

\appendix

\section{Testing the method} {\label{test}}

The main text described the material necessary to employ the maximum entropy inversion in any circumstance. 
The following points need to be noted 
(these are just tried and tested facts about the algorithm, many of which 
are illustrated here for the case of a toy problem shown in Fig(\ref{toy_prob}), whose solution is given in Fig (\ref{ini_guess})):

\begin{figure} [htbp]
   \begin{center}
     \includegraphics[width=0.5\textwidth,angle = -90]{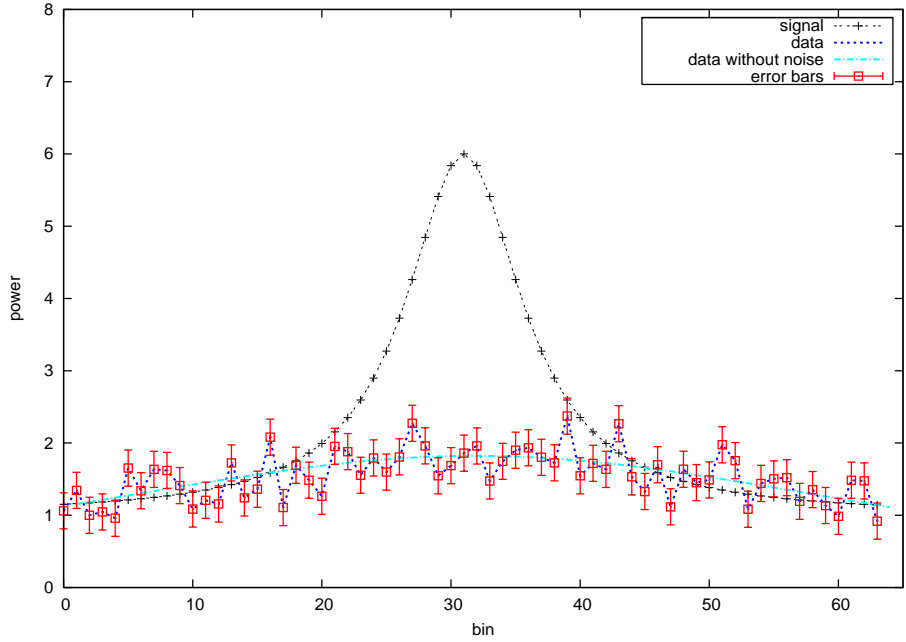}
   \end{center}
\caption{A toy problem to test the algorithm. The signal, which has a bump gets completely smoothed after the application of the 
kernel (chosen to be a Lorenzian profile), a known amount of random noise is then added giving the final data. Fig (\ref{ini_guess})
illustrates the recovery with two distinct initial guesses.}
  \label{toy_prob}
\end{figure} 

\begin{figure} [htbp]
  \begin{center}
    \includegraphics[width=0.5\textwidth,angle = -90]{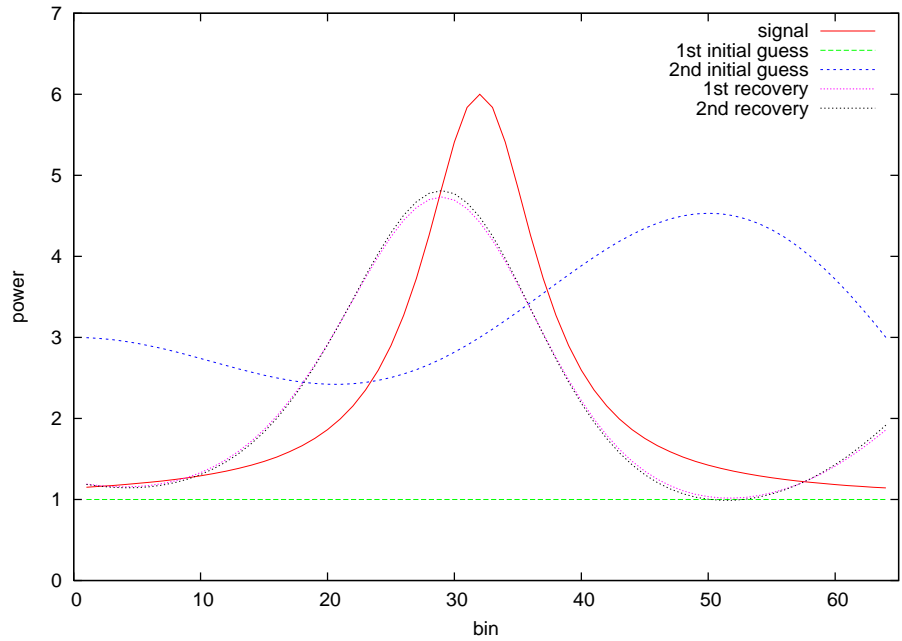}
  \end{center}
\caption{An illustration of the fact that even completely different initial guesses lead to the same final recovery 
(done for the toy problem of Fig \ref{toy_prob}). Notice that the location of the recovered bump and its amplitude are 
not exactly right: the quality of the recovery depends on many factors including the form of the kernel matrix itself.}
  \label{ini_guess}
\end{figure}

\begin{figure} [h]
  \begin{center}
    \includegraphics[width=0.75\textwidth,angle = 0]{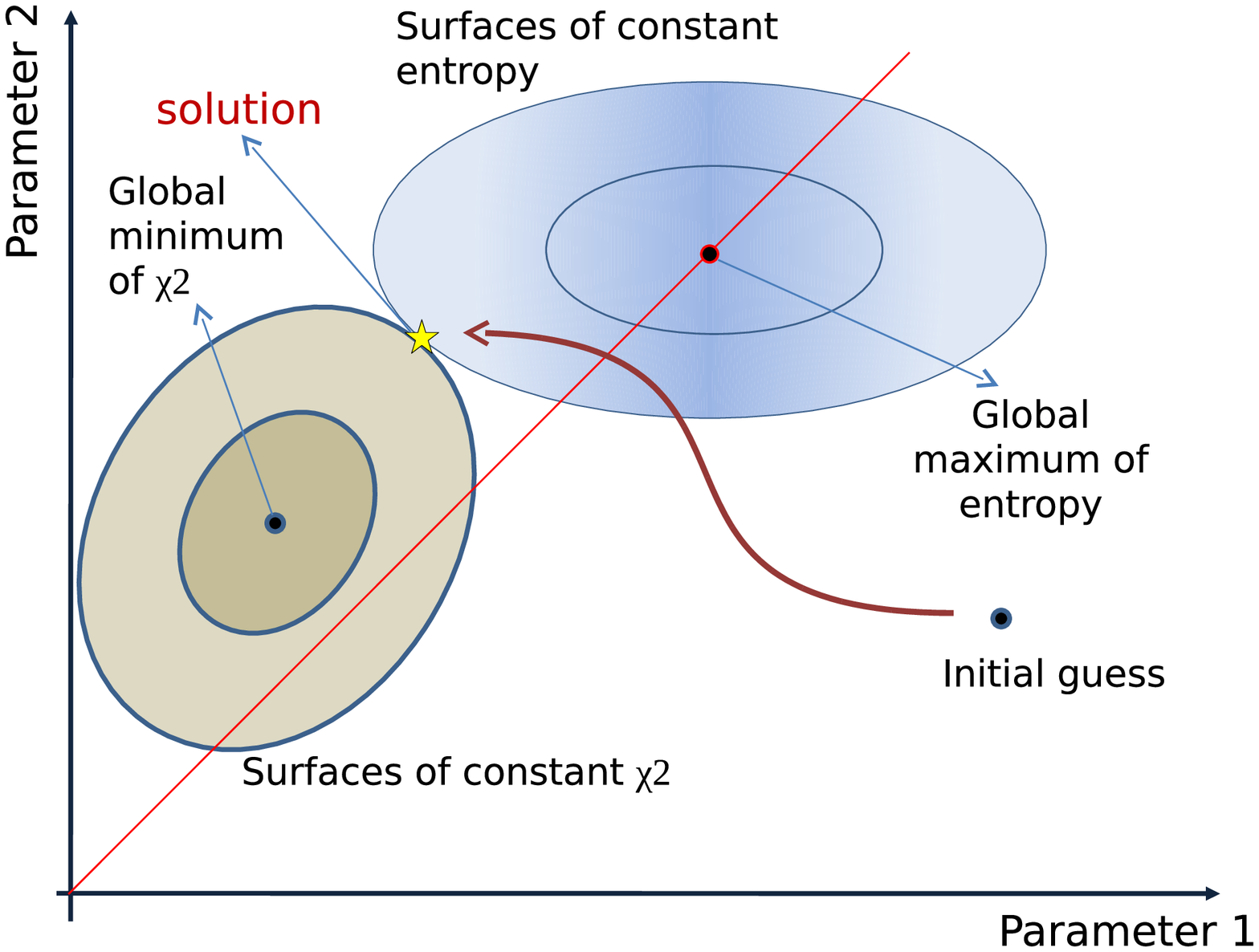}
  \end{center}
\caption{The red line in this Fig is $f_k = A$ line for various values of $A$. Changing the value of $A$ shall change the 
recovery because the point on the $\chi^2 = $constant surface with maximum entropy changes in the process.}
\label{effec_A}
\end{figure}

\begin{itemize}
 \item If we did not have any data available, the optimization problem would have involved maximizing entropy subject to no 
constraints. In such a scenario, the solution we should get must be $f_i = A$ as that is where the global maximum of entropy is.
 \item If the value of $A$ is such that the $\chi^2$ of the global maximum of entropy is smaller than $C_{\rm{aim}}$, then 
$f_i = A$ is itself the desired solution since ``the data are too noisy for any information to be extracted''
(see the last paragraph of page 113 of \cite{SB1984}).
 \item It is not a surprise at all that choosing too small value of $A$ should lead to negative value for entropy (the fact that 
depending upon the choice of $A$, sometimes we could be at locations in the parameter space with negative value of $S$ has no 
impact on the solution of the problem), see Fig (\ref{effec_A}). 
 \item In  Eq (\ref{Qeq}), $\alpha = \infty$ corresponds to the unconstrained maximization of $\tilde{S}$ irrespective 
of $\tilde{C}$. If we are too close to the global maximum of entropy ($f_i = A$), the value of $\alpha$ required to solve the 
constrained optimization problem in the subspace (for $\tilde{Q}$ defined by Eq (\ref{Qeq})) shall become too large. 
In this situation, it may be difficult to numerically find any solution for $\alpha$.
 \item As long as we do not stay too close to the global maximum of entropy (so that numerical problems such as those stated in 
the previous point above do not turn up), the choice of the initial guess for running the algorithm is immaterial. That is, 
all the initial guesses shall lead to the same answer (see Fig (\ref{ini_guess})).
 \item All the above problems can be easily avoided if we just choose a value of $A$ s.t. the $\chi^2$ of $f_i = A$ configuration
is much higher than the $\chi^2$ of initial guess (which better be more than $C_{\rm{aim}}$). Notice that this is not a 
requirement, just a trick. Also, this does not help us in finding any unique preferable value of $A$.
 \item For many kernels the exact value of $C_{\rm{aim}}$ chosen does not matter as far as the recovered $f$ is concerned, as long as 
the final value of $\theta$ becomes sufficiently small compared to a unit radian, all recoveries with different final 
$\chi^2$ are almost the same. The $\chi^2$ of signal (for a given realization) shall just fluctuate around (roughly) 
$n_d$, we have tested that if $C_{\rm{aim}}$ is set equal to $C_{\rm signal}$, the recovery does not change. This happens to be 
true e.g. for the case of CMB kernel, the case of our interest.
 \item The exact details of the shape of the final recovered solution does depend upon the actual value of $A$ chosen:  
the $\chi^2 = C_{\rm{aim}}$ surface can be thought of as a closed ellipsoidal surface in the $n_s$ dimensional $f$-space while 
as we change $A$, we define the line $f_i = A$ as being the location of global maximum of entropy for these different values of 
$A$. This will of course mean that as we change $A$, the place where the entropy is maximum on the $\chi^2 = C_{\rm{aim}}$ 
surface shall also change. Thus, as we continuously change $A$, we shall get a family of recoveries (see fig (\ref{effec_A})). 
So, the details of the recovered answer depends on the chosen value of this free (or adjustable) parameter.
But since the global maximum of entropy is at $f_i = A$, the value of $A$ represents the ``background'' (i.e.a priori) knowledge 
of how much the power in various bins is, without using any knowledge of data at all.  
 \item Whether the recovery is good or bad, depends on the details of the kernel. For the case of CMB kernel, we have tested that 
the recovery is often quite good.
\end{itemize}

\vspace{1in}

\noindent {\bf Acknowledgment:} 
The authors acknowledge the use of WMAP data and the use of codes such as CMBFAST, gTfast and CAMB.
The authors would also like to thank Tarun Souradeep (IUCAA, Pune) for reading 
through the manuscript and giving useful comments.
GG thanks Rajaram Nityananda (NCRA, Pune), Mihir Arjunwadkar (CMS, Pune university, Pune), 
Abhilash Mishra (CALTECH, Pasadena) and Ranjeev Misra (IUCAA, Pune) for discussions at various 
stages of the work. 
GG thanks Council of Scientific and Industrial Research (CSIR), India, for 
the research grant award No. 10-2(5)/2006(ii)-EU II. JP acknowledge support from the 
Swarnajayanti Fellowship, DST, India (awarded to Prof. Tarun Souradeep, IUCAA, Pune, India).
 

\begin{thebibliography}{99}

\bibitem{WMAP}
E. Komatsu et al. 2011 ApJS {\bf192} 18 (arXiv:1001.4538);
arXiv: 1212.5226; 

\bibitem{inflation} 
A.A. Starobinsky, Phys. Lett. B {\bf 91}, 99 (1980);
D. Kazanas, Ap. J. {\bf 241}, L59 (1980);
A. H. Guth, Phys. Rev. D {\bf 23}, 347 (1981);
A. D. Linde, Phys. Lett. {\bf B108}, 389 (1982);
A. Albrecht and P. J. Steinhardt,
Phys. Rev. Lett. {\bf 48}, 1220 (1982).

\bibitem{fluctuations}
A.A. Starobinsky, JETP Lett. 30, 682 (1979); 
Mukhanov V. F., Chibisov G. V., 1981, ZhETF Pis ma Redaktsiiu, 33, 549;
Hawking S. W., 1982, Physics Letters B, 115, 295;
A.A. Starobinsky, Phys. Lett. B 117, 175 (1982); 
Guth A. H., Pi~ S.Y., 1982, Physical Review Letters, 49, 1110.


\bibitem{infrev}
  A. Linde, arXiv: hep-th/ 0503203;
  D. H. Lyth and A. R. Liddle,
  \emph{The Primordial Density Perturbation}, Cambridge University Press, 2009;
  D. Baumann,
  arXiv:astro-ph/0907.5424v1;
  D. Langlois,
  arXiv:astro-ph/1001.5259v1;
  L. Sriramkumar,
  arXiv:astro-ph/0904.4584v1.

\bibitem{cmbf}
U. Seljak and M. Zladarriaga, Astrophys. J. \textbf{469}, 437 (1996).

\bibitem{camb}
A. Lewis, A. Challinor and A. Lasenby, Astrophys. J. 538, 473 (2000),
(http://camb.info/).


\bibitem{gtf}
Komatsu and Spergel, PRD 63, 063002 (2001); 

\bibitem{COSMOMC}
 Antony Lewis and Sarah Bridle Phys. Rev. D 66, 103511 (2002).

\bibitem{JPPSO}
 Jayanti Prasad and Tarun Souradeep 
 Phys. Rev. D {\bf 85}, 123008 (2012);
 Kiyotomo Ichiki1 and Ryo Nagata
 Phys. Rev. D 80, 083002 (2009);
 Kiyotomo Ichiki1, Ryo Nagata1, and Jun’ichi Yokoyama1
 Phys. Rev. D 81, 083010 (2010);

\bibitem{DKH}
Dhiraj Kumar Hazra et al JCAP10(2010)008;
Dhiraj Kumar Hazra et al 1106.2798v2;

\bibitem{PI1}
  Rajeev Kumar Jain, Pravabati Chingangbam, Jinn-Ouk Gong, 
L. Sriramkumar, Tarun Souradeep,
  JCAP 09 01: 009, 2009
  [arXiv:astro-ph/0809.3915].

\bibitem{PI2}
  Rajeev Kumar Jain, Pravabati Chingangbam, L. Sriramkumar, Tarun Souradeep
  Phy. Rev. D {\bf 82} 023509 (2010)
  [arXiv:astro-ph/0904.2518].




\bibitem{cusps}
  Ryo Saito, Jun'ichi Yokoyama, Ryo Nagata JCAP06(2008)024;
  Rajeev Kumar Jain, Pravabati Chingangbam, L. Sriramkumar
  JCAP 07 10: 003, 2007. [arXiv:astro-ph/0703762];
  Sirichai Chongchitnan, George Efstathiou JCAP 07 01: 011, 2007.

\bibitem{nonTevol2}
  Edgar Bugaev, Peter Klimai
  Phys. Rev. D 78, 063515 (2008).

\bibitem{Leach1}
  Samuel M. Leach and Andrew R. Liddle  
  Phys. Rev. D 63, 043508 (2001)
  [arXiv:astro-ph/0010082].

\bibitem{Staro1992}
  A. A. Starobinsky
  Pis'ma Zh. \'{E}ksp. Teor. Fiz. 55, 477 (1992)
  [JETP Lett. 55, 489 (1992)].



\bibitem{ppsfeatures}
  A. Shafieloo and T. Souradeep
  Phy. Rev. D {\bf 70}, 043523 (2004);
  R. Sinha and T. Souradeep
  Phy. Rev. D {\bf 74}, 043518 (2006);
  A. Shafieloo, T. Souradeep, P. Manimaran, P.K. Panigrahi and R. Rangarajan
  Phy. Rev. D {\bf 75}, 123502 (2007);
  A. Shafieloo and T. Souradeep
  Phy. Rev. D {\bf 78}, 023511 (2008);  
 Gavin Nicholson and Carlo R. Contaldi JCAP07 (2009) 011;
  Tocchini-Valentini, D., Hoffman, Y. and Silk, J.
  MNRAS, 367: 1095-1102, 2006.

\bibitem{armantarundec09}
  J. Hamann, A. Shafieloo and T. Souradeep
  JCAP 10 04: 010,2010.

\bibitem{Dodelson}
S. Dodelson, \textit{Modern Cosmology} (Academic Press, San Diego, U.S.A., 2003).

\bibitem{nr}
Press, William H.; Teukolsky, Saul A.; Vetterling, William T.; Flannery, Brian P., 
\textit{Numerical recipes in FORTRAN. The art of scientific computing}
Cambridge: University Press, 1992, 2nd ed.

\bibitem{SB1984}
J. Skilling and R.K. Bryan
Mon. Not. R. Astr. Soc. (1984) \textbf{211}, 111 - 124.

\bibitem{SB_others} 
Skilling, J., and Gull, S.F. 1985, inMaximum-Entropy and Bayesian Methods in Inverse Problems,
C.R. Smith and W.T. Grandy, Jr., eds. (Dordrecht: Reidel). 
Skilling, J. 1986, in Maximum Entropy and Bayesian Methods in Applied Statistics, J.H. Justice,
ed. (Cambridge: Cambridge University Press).
Gull, S.F. 1989, in Maximum Entropy and Bayesian Methods, J. Skilling, ed. (Boston: Kluwer).


\bibitem{matrixcomp}
Matrix Computations (Third Edition) by Gene H.Golub and Charles F.Van Loan




 
\end{thebibliography}
\end{document}